\documentstyle[12pt,aaspp4]{article}

\lefthead{Santos Jr. \& Frogel}
\righthead{Stochastic Effects on Star Cluster Colors}

\begin{document}

\title{Integrated Near-Infrared Colors of Star Clusters:\\
       Analysis of the Stochastic Effects on the IMF}

\author{Jo\~ao F. C. Santos Jr.\altaffilmark{1} and 
        Jay A. Frogel\altaffilmark{2}}
\affil{Department of Astronomy, The Ohio State University,
        174 W. 18th Avenue, Columbus, OH 43210, USA}

\altaffiltext{1}{CNPq fellow -  Brazil}
\altaffiltext{2}{Also at Department of Physics, University of Durham, England}

\begin{abstract}
In this paper we examine the influence of stochastic effects on 
the integrated near-infrared light of star clusters with ages
between $7.5<\log{t}<9.25$. To do this, we use stellar evolution 
models and a Monte Carlo technique to simulate the effects of 
stochastic variations in the numbers of main sequence, giant, 
and supergiant stars for single-generation stellar
populations (SSPs). The fluctuations in the
integrated light produced by such variations are
evaluated for the $VJHK$ bands. We show that the $VJHK$ light of
the star clusters can be strongly affected by plausible 
stochastic fluctuations in the numbers of bright but scarce
stars. In particular, the inclusion of thermally pulsing AGB
stars in the stellar evolution models yields integrated colors
with values in agreement with the spread seen for Large Magellanic
Cloud clusters that are known to have significant number of 
AGB stars.
Implications of this analysis
are important for studies of the integrated light of stellar
populations where it is not possible to resolve individual stars.
\end{abstract}


\section{Introduction}

The main question addressed in this work is 
how large is the contribution of small numbers of bright stars to the
near-infrared (near-IR) integrated light of a star cluster? 
For a cluster of given age and metallicity, the size of the 
contribution will depend on
the evolutionary state of the stars in question as 
characterized by their temperature and luminosity.
More generally, the effects of changing small numbers of stars will 
depend on the total mass of the cluster and the cluster's
initial mass function, its IMF. For stochastic variations of a 
given size the integrated light from less 
massive clusters will obviously be more strongly influenced 
than will the light from more massive clusters. 
The advantages of including the $JHK$ bands in the analysis is that
the effects of variations in the numbers of later type giants and
supergiants in a cluster on the integrated light
will be much more evident than if only optical wavelengths were examined.

The age range analyzed $(7.5\leq\log{t}\leq9.25)$ was chosen
because the available isochrones contain all relevant stellar
evolutionary phases up to the end of the C-burning phase for 
stars with ${\rm M}>7{\rm M}_{\odot}$ and up to the end of the
early asymptotic giant branch (E-AGB) for stars with 
$2\leq{\rm M}/{\rm M}_{\odot}\leq5$; for stars with 
${\rm M}\leq1.7{\rm M}_{\odot}$, corresponding to  $\log{t}>9.25$,
the evolution is followed up to the He-flash (\cite{sch92}).
Consequently, although the thermally pulsing asymptotic giant 
branch (TP-AGB) for intermediate mass models is missing, the
He-burning phases are covered by the models in the age range
considered. The TP-AGB according to models by \cite{bed88} was
taken into account in a later stage of this work providing 
information on its contribution to the integrated light of stellar
populations and allowing a better approximation to real star
clusters concerning their integrated colors and stochastic
fluctuations. The lower
age limit selected avoids massive stars in evolutionary stages
with very high mass loss rates, for which bolometric correction
and temperatures are very uncertain.
The present work is restricted to solar metallicity since
the bolometric corrections and the color-temperature relations
available come from solar neighborhood stars.
This restriction, though, do not affect our primary goal which 
is to examine stochastic effects in stellar populations on the
integrated colors of those populations.

Similar studies using visible bands have been carried out by, e. g.,
\cite{bar77,bec83,chi88,gir93,gir95}.
In particular, \cite{chi88}
employ a homogeneous set of stellar evolution models that include 
convective overshooting and cover the whole range of evolutionary phases.
They calculated integrated ($U-B$) and ($B-V$) colors and synthetic H-R 
diagrams of star clusters as a function of age, and applied their results
to the known age and color distribution of clusters in the Large 
Magellanic Cloud (LMC). Beside the stochastic effects on
the IMF, they also take into account the effects of dispersion in the 
stellar ages.  In summary, among other important results, their 
analysis indicates that integrated $UBV$ colors show a significant 
spread for ages at which few luminous AGB stars are present. 
Thus, they conclude that it is necessary to take stochastic effects 
into account when the integrated light of a cluster is used to
estimate the cluster's age and chemical composition.
\cite{gir95} employ the photometric models of simple stellar
populations (SSPs) by \cite{ber94} and the integrated $UBV$
colors of LMC star clusters (\cite{bic96}) to investigate the
history of star formation in the LMC. They present theoretical integrated
$UBV$ colors of SSPs with ranges in age and metallicity that span 
the observed ranges for the clusters.  They analyze the dispersion in 
the observed integrated colors of the clusters and conclude that in $UBV$
stochastic effects dominate the color dispersion of the young clusters.
Dispersions in the colors due to variations in reddening and 
metallicity 
become significant for clusters older than 300\,Myr. In summary, 
the highest dispersion in the integrated colors of the SSPs occurs 
for ages between 6.3 and 25\,Myr as a consequence of the number 
and luminosity of red supergiants. Similar results were obtained 
for different metallicities, except for smaller dispersions in 
young, metal poor SSPs where red supergiants are rare compared 
to a more metal rich population. 
  
The near-IR light of a stellar system can be dominated 
by only a small number of luminous, 
red stars most of whose energy is emitted in the near-IR (e. g. 
\cite{per83,fro90}). Therefore,
stochastical variations in the numbers of these stars  
could lead to erroneous conclusions concerning the age and/or
metallicity of the system if such estimates are based on integrated
light measurements. 
Thus, in the present study we analyze the effect on the
integrated $JHK$ light from a single-generation
stellar population, of stochastic variations in its IMF.
This issue is addressed here in the hope of 
setting useful constraints on stellar populations of galaxies, allowing
one to decide how much an integrated observation in $JHK$ of a star system
of certain age and metallicity can be affected
by the stochastic nature of the IMF in single-generation stellar
populations.

In section~2 the data employed and the procedure carried out to transform 
the theoretical H-R diagram into observational color-magnitude diagrams
are described. Section~3 presents the Monte Carlo technique used to
carry out the simulations and a comparison between a young SSP and an
intermediate age one. Section~4 discusses the relationship between the stochastic effects and
the age of a SSP. In section~5 the results are compared with observations
of LMC star clusters.
Finally, section~6 presents the conclusions. 

\section{Data}

\subsection{Stellar Evolution Models}

The stellar evolution models by Schaller et al. (1992, hereafter
SSMM) were employed 
in the present study. Basically they present
models from 0.8 to 120 ${\rm M}_{\odot}$, at metallicities $Z=0.001$
and solar, with mass loss and convective overshooting. The 
stellar tracks were
transformed to isochrones by using the code made accessible by G. 
Meynet. The fainter end of the main sequence (MS) ($m<0.8{\rm M}_{\odot}$)
was extended down to 0.07${\rm M}_{\odot}$ by using the information in 
\cite{bes91} ($T_{\rm eff}$ and $L$) and \cite{van83}
(mass-luminosity relation to get $m$). 
For most of the ages investigated less massive stars have not
yet contracted to the MS phase, as at least 1\,Gyr is needed for a 
complete MS to be formed (\cite{mer96}, and references therein).
However, a complete MS was used here, even for the younger SSPs, as
a rough approximation to real clusters where low mass stars are in
pre-MS phases. This approximation should be a lower limit in terms of
near-IR integrated light contribution, since low mass contracting stars
are cooler and brighter than their future MS location. 

In the present context, the fact that
the models do not take into account evolution beyond the E-AGB for
intermediate mass stars must be considered. \cite{fer95} carried out 
$BVJHK$ observations of stars in 12 intermediate age LMC clusters
showing that around 60\% of their integrated $K$ light 
comes from AGB stars, consistent with the results of Persson et al. (1983)
and Frogel et al. (1990). \cite{cha91}  constructed stellar
population models by using \cite{mae89} stellar 
evolution models complemented with observations of TP-AGB. Their 
models indicate that the whole AGB contribute with 
60\% at J and 70\% at K to the integrated light of a 0.4-2\,Gyr SSP, 
the TP-AGB stars alone contributing with 20\% at K 
and a little less at J. The agreement between these studies 
and their results support the conclusion that the E-AGB is the
major contributor to the whole AGB in the integrated near-IR
light of intermediate age star clusters. The TP-AGB phase, though,
has not negligible contribution and should be kept in mind,
especially when stochastic variations in such stellar populations
are to be evaluated. 

\subsection{Near-Infrared Colors}

The $JHK$ (and $V$, in the Kron-Cousins system) colors, in a 
standardized system, are from \cite{bes88} for MS stars from B8 to 
K5 and for giants from
G0 to M6, and from Bessell (1991, hereafter B91) for later MS 
spectral types. When needed, the colors for MS stars earlier than 
B8 were linearly extrapolated using the hottest spectral types 
available; the same has been done for giants earlier than G0.
Colors for O5-M5 supergiants, all considered as evolved stars
with  $M_{\rm bol}\leq-4$, are from \cite{koo83}. In this case,
the transformations in \cite{bes88} were used to bring
Koornneef's colors into their system.
 
\subsection{Bolometric Corrections and Temperature Scale}

In order to transform the theoretical H-R diagram into a color-magnitude
diagram (CMD) for comparison with observations,
bolometric corrections and temperature scale compiled
by Schmidt-Kaler (1982, hereafter SK82) for giants, supergiants and
for MS stars of spectral type earlier than M and those determined 
by B91 for fainter stars were employed.
The link between the colors and BC$_{V}$ and $T_{\rm eff}$ was 
guided by the spectral type provided in both sets of data. An 
exception to this procedure was adopted for the M dwarfs, for which
B91 gives colors, $T_{\rm eff}$, and all other information  needed 
to get BC$_{V}$ (his Table 2).

The bolometric correction was defined here as BC$_{V}=M_{\rm bol}-M_{V}$
and its zero-point was set by adopting
BC$_{V\odot}=-0.08$ ($M_{\rm bol\odot}=4.75$). 
The consistency between B91 and SK82 data was checked
by looking at the flux density of a zero-magnitude star in both scales.
The $V$ magnitude difference at $V=0$ is 0.03 mag brighter in
the Johnson system (adopted by SK82), which is insignificant
given the uncertainties.
The temperature scale adopted was checked at spectral type K7V (where
the data from the different sources were merged) and no significant
discrepancy was found. On the other hand, at this spectral type there
was a difference of 0.09 mag in the bolometric correction of SK82 and
B91 (in the sense that BC is larger in SK82). The option adopted was 
to smooth out the BC around  $T_{\rm eff}=4000$\,{\rm K} (i.e. K7V) 
as long as this procedure is confined to the fainter end of the MS 
and will not have a major effect on integrated light models.
   
\section{Simulations of Single-Generation Stellar Populations}

\cite{bar77,bec83,chi88,gir93} and \cite{gir95} investigated 
the importance of stochastic effects on the IMF of star clusters in 
the $UBV$ bands.
Similarly, in the present study we used a Monte Carlo technique 
as an approach to assess the impact of stochastic variations in the 
IMF of single-generation systems on near-IR bands.

The Monte Carlo integration used simulates the stellar mass
distribution of
a system of a given age: a randomly generated mass is weighted on
the IMF and the process is repeated until the number of stars sampled
reachs a preset value.  A total of 100 simulations of systems 
containing a variable number of stars ($10^2$ to $10^6$) is 
performed and the mean and dispersion of the integrated colors are computed.

Following is a more detailed description of the procedure used to carry
out the simulations.  An introduction to Monte Carlo methods is given
by, e. g., \cite{all96} and \cite{pre92}.

\subsection{Theoretical Background}

 The IMF of a stellar system gives the relative number of stars 
within a mass range and it is employed in its usual definition 
\begin{equation}
\label{imf}
                  \Phi(m)=dN/dm=Am^{-(1+x)},
\end{equation}

where $A$ is a constant and $x$ the IMF slope. $x=1.35$ is the value
corresponding to the \cite{sal55} luminosity function.
$A$ is determined by integrating the IMF from the lowest ($m_l$=0.07) 
to the highest stellar mass ($m_u$=120) in the system 
and by normalizing the number of stars in the system to 1. Specifically,
\begin{equation}
        A=\frac{x}{m_l^{-x}-m_u^{-x}}
\end{equation}
This normalization is critical for the present application of the
Monte Carlo method since now $\Phi(m)$ can be treated as a probability 
distribution function (PDF) which gives the probability that a random
mass $m'$ is in the range between $m$ and $m+dm$.  $\Phi(m)$ now obeys
the following conditions:
\begin{equation}
   \Phi(m)\ge0, \mbox{    }{\rm for} \mbox{   } {\rm any} \mbox{    } m 
\end{equation}
\begin{equation}
   \int_{m_l}^{m_u}\Phi(m')dm'=1
\end{equation}
where $m_l$ and $m_u$ are the lower and upper mass limits of the stellar
system. 

Now, the PDF $\Phi(m)$ can be transformed into a new PDF 
$g(N)$, where $N$ is a single-valued function of $m$.
Therefore, the probability of occurrence of the random
variable $N'$ within $dN$ and the probability of occurrence of the random
variable $m'$ within $dm$ must be the same, i. e.,
\begin{equation}
\label{prob}
       |\Phi(m)dm|=|g(N)dN|
\end{equation}

From Eq. (\ref{imf}), a cumulative distribution 
function (CDF) can be constructed which gives the probability 
that the mass $m'$ is less than or equal to $m$, i. e.,
\begin{equation}
\label{cdf}
           N(m)=\int_{m_l}^m\Phi(m')dm'
\end{equation}
and, consequently, $dN/dm=\Phi(m)$, resulting from Eq. (\ref{prob})
\begin{equation}
        g(N)=1, \mbox{      } 0 \leq N \leq1
\end{equation}

This means that the CDF is a uniform function for which any value 
is equally likely in the interval [0,1].

The CDF property of uniformity between 0 and 1 
allows one to use a random number generator (namely, that of Parker 
and Miller with Bays-Durham shuffle as presented by \cite{pre92}) 
to sample values for the CDF $N(m)$.
Since the random variable $m$ and the CDF $N(m)$ are related uniquely, 
it is possible to sample $m$ by sampling $N$ first and then to solve 
the result of the integration of Eq. (\ref{cdf}) for $m$, i. e.,

\begin{equation}
      m=[m_l^{-x}(1-N)+m_u^{-x}N]^{-1/x}
\end{equation} 
Each stellar mass $m$ generated in this way is interpolated linearly 
on an 
isochrone in order to get the corresponding $\log{T}_{\rm eff}$ and
$\log{L}$. The correlations assembled in section 2.3 were then linearly
interpolated to the colors corresponding to these values of 
$\log{T}_{\rm eff}$ and $\log{L}$. 
The mass difference between consecutive points on an isochrone is 
proportional to the evolutionary lifetime. Therefore a random mass
will have larger probability to be sampled in evolutionary stages
for which there is a relatively wider range of masses indicating
longer lifetimes. Consequently, for a SSP of a given age, the fact
that the number of stars in each evolutionary phase is proportional
to its lifetime is taken into account.
Integrated fluxes and colors are then computed 
by appropriately combining the fluxes and colors from the individual stars
with the weights as described above.

\subsection{A Comparison Between Young and Intermediate Age SSPs}
 
In this section, a comparison of the simulations carried out for solar
metallicity SSPs with ages 32\,Myr ($\log{t}=7.5$) and 1\,Gyr 
($\log{t}=9.0$) is presented.
With the adopted criterion that stars with $M_{\rm bol}\leq-4$ are
supergiants, it turns out that the post MS stars in the 32\,Myr SSP 
are slightly brighter than this limit.  We also computed simulations 
treating these stars as giants for comparison. 
Although the colors of the giants and supergiants differ, 
we find no significant differences between these two sets of 
simulations apart from those resulting from the stochastic effects.

Fig.~1 shows how the young stellar population is distributed in
CMDs formed by both visible and near-IR bands (a,c) and exclusively
by near-IR bands (b,d). Figs. 1a and 1b show different CMDs for the
same SSP simulated, while Figs. 1c and 1d show the same for another
simulation. Each SSP has  $10^5$ stars and an IMF with $x=1.35$.  Only
the upper MS and the supergiants, the only regions for which 
the Monte Carlo simulations reveal significant variations, are shown.
The integrated colors corresponding to each CMD are indicated on the 
panels. Fig. 2 corresponds to Fig. 1 for an intermediate age
stellar population. 
We also ran an experiment to see, in the absence of any stochastic 
effects, how many stars would be required to produce at least one 
post MS star in such SSPs.  We found that young SSPs ($\log{t}=7.5$)
containing 140 stars 
for an IMF slope $x=-0.5$, 8990 stars for $x=1.35$ and 136540 stars 
for $x=2.0$ should contain one post MS star if a continuous IMF 
is considered. Similarly, intermediate age SSPs ($\log{t}=9.0$)
containing 170 stars 
for an IMF slope $x=-0.5$, 890 stars for $x=1.35$ and 5590 stars 
for $x=2.0$ should contain one post MS star.
For SSPs less populous than these the stochastic 
nature of the IMF can still produce post MS stars and, consequently, 
color variations as large as those found for SSPs with number of
post MS Npms=1. If 
an SSP is far less populous then there would be no post 
MS stars, and the integrated fluxes would be dominated by upper 
MS stars with little spread in color, which results in smaller 
color fluctuations.

\placefigure{f1} 
\placefigure{f2}

Fig. 3 presents the integrated colors ($V-K$), ($J-K$), ($H-K$) 
as a function of the number of post MS stars for the young SSP
(a, b and c) and the older one (d, e and f), where
3 slopes of the IMF are also considered. Fluctuations in the colors
come from 100 simulations for SSPs of fixed total number of stars and
are represented by twice the standard deviation from the mean color.
($V-K$) has the largest fluctuations because it is most sensitive to 
the slope of the spectral energy distribution of the stars in the 
system. 
The fluctuations are also larger the fewer post MS stars there
are in the system.  As their number increases, not only do the 
color fluctuations decrease but the mean colors get redder as well.
As may be seen from Fig. 3, the stochastic effects are always 
more significant than variations due to changes in the IMF slope 
for the range studied here except for the young SSP when the number 
of post MS stars 
gets to be 100 -- then ($V-K$) is larger for $x=-0.5$ than it is 
for steeper values of the IMF slope.

\placefigure{f3} 

Relative flux contributions from MS and post MS stars 
for well-populated SSPs with 100 post
MS stars are plotted for the 3 slopes of the IMF in Fig. 4. 
For the young SSP (left side of the Fig. 4) the $V$
flux contribution is distributed between the MS turnoff (B type
stars) and the
supergiantes, with insignificant contribution coming from the lower MS.
As $x$ increases from -0.5 to 2, the MS turnoff $V$ contribution 
increases from 40\% to 60\%, while the supergiants one decreases
from 60\% to 30\%.
The supergiants dominate the integrated $JHK$ bands for all IMF slopes,
followed by the MS turnoff, which contributes, at most, with 15\% of
the near-IR light. For the older SSP (right side of Fig. 4)
the $V$ flux contribution is distributed again between the MS turnoff
(types A and F) and the post MS stars. As $x$ increases from -0.5
to 2, the turnoff (A+F stars) $V$ contribution increases from 53\%
to 65\%, while the giants one decreases from 45\% to 30\%. 
Differently from the young SSP, there is a significant contribution
in the near-IR bands from the turnoff: as $x$ increases from
-0.5 to 2,  A and F stars together increase their contribution to
$H$ and $K$ from 20\% to 27\%, the contribution in $J$ reaching 
40\%. The giants contribute with 80\% ($x=-0.5$) to 60\% ($x=2$)
in  $H$ and $K$ and with 72\% ($x=-0.5$) to 53\% ($x=2$) in $J$.
We can compare these results with that found for an old 
($\approx$\,10\,Gyr) stellar population (e.g. \cite{fro88}). 
For such a population, evolving dwarfs, subgiants  and K giants 
dominate the $B$ and $V$ integrated light, regardless of the IMF. In $JHK$  
the M giants are the largest single contributors to the flux. 
For $x=1.35$ their contribution reaches
50\% of the total $K$ band light and is greater still for
$x=0$. In connection with the present study, Frogel's models show 
the relative importance of red giants in the integrated light
of composite systems that contain an older stellar component, 
and whose contribution can change from $V$ to $K$ by significant 
amount.

Subsequent analysis (section 5) shows that the inclusion of the
TP-AGB can influence significantly the integrated colors in the
case of the SSP with $\log{t}=9$.

\placefigure{f4} 

\section{SSP Age and the Stochastic Effects}

For the whole range of ages ($7.5<\log{t}<9.25$) SSPs were computed
in steps of $\Delta\log{t}=0.25$. The behavior of the integrated 
($J-K$) color
as a function of $\log{t}$ for SSPs with different total masses
and IMF slope $x=1.35$ is shown in Fig. 5. Panel (a) shows the color
evolution of SSPs with $10^3$ and $10^6$ stars; it is also shown
the percentage of post MS stars for each age (=Npms/(N.100)). 
When SSPs of identical total number of stars are compared, they can
be associated to the evolution of a single star cluster. 
However, young clusters with a small number of stars 
are more difficult to interpret in terms of evolution since their 
post MS may be  underrepresented, which produces large
fluctuations in the integrated colors.
Fig. 5a shows that:
\begin{itemize}
\item Less massive clusters with $10^3$ stars present a gradual
reddening as they age. 
\item Young massive clusters with $10^6$ stars can have integrated
(J-K) as red as those of older clusters.
\item Young star clusters have integrated near-IR colors 
very sensitive to their masses.
\item Older star clusters of different masses have small spread 
in color.
\end{itemize}

In Fig. 5b
the color evolution is presented for SSPs with a fixed number
of post MS stars. The main difference between systems with Npms=100
and Npms=500 is the smaller color fluctuations for the last ones,
the mean color having similar behavior to that of the more massive 
SSPs with N$=10^6$ stars (see Fig. 5a).

\placefigure{f5} 

Well-populated SSPs with at least
100 post MS stars have their integrated colors modulate by a 
combination of factors like the ratio between the number of upper
MS and post MS stars and the extension in temperature of the blue loop
and its luminosity compared to the asymptotic giant branch and to
the MS turnoff. 

For $x=1.35$, 
color fluctuations of 10\% ($1\sigma$ from the mean) are reached
when there are 70 post MS stars for $\log{t}=7.5$, 80 for 
$\log{t}=8.0$, 350 for $\log{t}=8.5$ and 40 for $\log{t}=9.0$. 
The larger the number of post MS stars, the smaller the color
fluctuations.
The extended E-AGB 
at age $\log{t}=8.5$ is the main feature responsible for the largest
fluctuations in the integrated color of that SSP compared to the
other ages. 

Fig. 6 presents the evolution of the $VJHK$ flux contribution of 
MS and  post MS stars to the integrated light of SSPs built with
$x=1.35$ and 100 post MS stars. In summary, the results indicated:
\begin{itemize}
\item In $V$, as the age increases from $\log{t}=7.5$ to 8.25 the MS B stars contribution gets between 55\% and 63\%, while the post MS
contribution decreases from 40\% to 25\%. The contribution of redder
MS spectral types being less than 15\%. 
\item For $\log{t}=8.5$ MS B, MS A
stars and post MS stars share most of the 
total integrated $V$ light, contributing with 40\%, 27\% and 23\%
respectively. The major contributors for the $V$ light of older SSPs
change dramatically: for $8.5<\log{t}\leq9.25$, the post MS 
contribution
increases from 25\% to 35\%, the MS A stars decrease strongly their 
contribution from 60\% to 0\% while the MS F ones increase theirs
from 10\% to 55\%.
\item In $H$ and $K$, the post MS contribution dominates throughly the
near-IR integrated light from $\approx$90\% (at $\log{t}=7.5$) to
$\approx$70\% (at $\log{t}=9.25$); the contribution of the turnoff
increases for the older SSPs reaching 25\% at most.
\item In $J$, the turnoff MS types have significant contribution
especially for the older SSPs: the B type peaks at $\log{t}=8.25$
with 23\%, the A type at $\log{t}=8.75$ with 32\% and the F type
at $\log{t}=9.25$ with 30\%.
\end{itemize}

\placefigure{f6} 

\section{Comparison with Observations}

We compare the integrated color models and their fluctuations 
to the integrated colors
of Large Magellanic Cloud clusters (from \cite{per83}) in
Fig. 7a. 
Regarding the age spread of the models computed, they are in 
principle adequate for a comparison with 
clusters of SWB types ranging from II to V. 
However, in most of  the cases,  the models are not expected
to encompass
clusters with SWB type IV and V because the stellar evolution
models do not take into account evolution  beyond the E-AGB 
for intermediate mass stars,
where bright AGBs like Carbon stars can contribute significantly
to the integrated light of a cluster.

\placefigure{f7} 

Frogel et al. (1990) analyzed the AGB contribution  to the integrated
light of Magellanic Cloud clusters. They show that Carbon stars
(originated by TP-AGB stars) are only found 
in clusters of SWB types IV-VI and account for 
50\% to 100\% of the bolometric luminosity from the AGB above the
RGB tip (He-flash). The same type of stars contribute
with 30\% to the bolometric luminosity of a cluster with age
around 2\,Gyr (SWB V-VI), the whole AGB ($M_{\rm bol}\leq-3.6$)
contributing with 40\%.

The agreement between the younger SSP models with
clusters of types SWB II and III in the color-color diagram of 
Fig. 7a is indicating that the isochrones are
essentially complete for ages around $\log{t}=7.5$. More importantly, 
the small scatter of clusters around the younger SSP models 
indicates that their integrated colors are compatible with SSPs
showing an almost continuous IMF, since big fluctuations in color
are not needed to explain the young cluster distribution.  
It seems  that the inclusion of the TP-AGB phase in the stellar
evolution models would improve the older SSP integrated colors,
matching the observations.

In order to address the aforementioned point, the influence of the
TP-AGB on the integrated colors was evaluated quantitatively by
complementing the isochrones with the upper AGB semi-empirical models
of \cite{bed88}, which consider that stars with initial masses in the
range $0.8<{\rm M}/{\rm M}_{\odot}<8$ will pass through that phase.
The BC$_{K}$~$vs$~($J-K$) and $M_{\rm bol}$~$vs$~$\log{T_{\rm eff}}$
relationships 
in \cite{fro80} together with the AGB tip near-IR colors for LMC
clusters observed by \cite{fer95} were used. The TP-AGB has a maximum
extension for the isochrone with $\log{t}=9$, for which it spans
$\Delta{M}_{\rm bol}=4.4$ and $\log{T_{\rm eff}}=0.22$. On the
observational side the corresponding values are $\Delta{K}=5.5$,
$\Delta$($V-K$)=3.2 and $\Delta$($J-K$)=1.3. 

Tables 1 to 8 list, for each age, integrated near-IR colors and their
respective fluctuations for three IMF slopes  after the AGB to have been
extended up to the end of the thermal pulses phase. SSPs with 
three values of the expected integrated V flux are presented and 
provide an useful tool for comparison with stellar systems of different
sizes.

Indeed, when the TP-AGB is taken into account, the near-IR 
integrated color models
fit satisfactorily the observed colors of LMC clusters as shown in
Figure 7b. In particular, 
the $\log{t}=9$ SSP with $F_{\rm V}=1000$ and IMF slope $x=1.35$ (Table 7)
give an excelent match to the locus of the reddest LMC clusters in Fig. 7b.

In order to check the sensitivity of the integrated colors on the bolometric
correction and temperature scale adopted, that information as provided 
by \cite{fro87} for the AGB in the Galactic bulge was used to build SSPs
of $\log{t}=9$. The integrated colors obtained at $\log{t}=9$ with the 
two different sets of theoretical to observational transformations 
indicated that all colors and fluctuations are smaller when the Galactic
bulge constraints were used. The larger the stellar system, the larger
the differences. For an SSP with  $F_{\rm V}=1000$ the color differences 
are 7\% in ($V-K$), 15\% in ($J-K$) and 26\% in ($H-K$); the color 
fluctuations differ by 18\% in ($V-K$), 38\% in ($J-K$) and 41\% in 
($H-K$).

\section{Concluding Remarks}

The present work indicated the importance of stochastic effects
on the near-IR integrated colors of young and intermediate age SSPs.
If integrated observations of galaxies can be represented by a mixture
of integrated colors of SSPs as those analyzed here, then it is 
clear that studies
of stellar populations in galaxies must take into account the influence
of bright and scarce stars on their integrated light, which can
affect significantly determinations of age and metallicity. 
Combining independent estimates (other than by integrated light observations)
of age and metallicity for a given galaxy with its integrated
colors would allow one, by using the present technique, to quantify
how important are the stochastic effects for that particular 
stellar population.

We show that the inclusion of the TP-AGB phase for intermediate 
mass stars in the stellar evolution models by means of the
semi-empirical approach presented here brought the 
integrated
near-IR colors of intermediate age SSPs in excelent agreement with
the spread of observed colors of LMC
clusters as far as the stochastic variations are considered.
This result reflect the expected fact that the mean integrated 
colors are redder and subject to higher fluctuations than those
generated from stellar evolution models without the TP-AGB phase.

In general, the main conclusions are:

\begin{itemize}
\item Near-infrared bands combined to the optical $V$ band are
useful tools to the investigation of the red stellar population
and its stochastic fluctuation in star clusters and, consequently,
in more complex systems like galaxies.
\item The $VJHK$ integrated colors of SSPs, when compared to those
of LMC star clusters,  gave information on the importance of
taking into account  TP-AGBs
like Carbon stars in the stellar evolution models.
\item The importance of stochastic fluctuations in the integrated
$VJHK$ light of star clusters has been confirmed in this analysis,
corroborating studies in the visible spectral range; specifically,
general
properties derived from the integrated light of composite stellar
populations must account for the influence of stars undergoing fast 
yet bright evolutionary stages.
\end{itemize}

\acknowledgments

JAF acknowledges NSF grant No. AST92-18281. He also thanks 
Roger Davies
for his hospitality and PPARC for their partial support while a
Visiting Senior Research Fellow at Durham University.
JFCSJ thanks a 
postdoctoral fellowship from the Brazilian Institution CNPq.
We also thank Marc Pinsonneault for interesting remarks and the
referee for some good suggestions.

\newpage
\figcaption[fig1.ps]{Two simulations of a young SSP with the same 
parameters
are shown in order to exemplify the stochastic effects in different
CMDs. Both simulations represent
SSPs having $10^5$ stars and IMF slope $x=1.35$. Their corresponding
integrated colors are indicated.
Only stars more massive than 1.1\,M$_{\odot}$ are plotted. \label{f1}}

\figcaption[fig2.ps]{Same as Fig. 1 for an older SSP. \label{f2}}

\figcaption[fig3.ps]{The behavior of the near-IR integrated colors 
and their 
fluctuation according to the number of expected post MS stars of SSPs 
with $\log{t}=7.5$ (a, b, c) and $\log{t}=9.0$ (d, e, f) are shown. 
Different IMF slopes are considered.  \label{f3}}

\figcaption[fig4.ps]{The relative flux contribution of MS and 
post MS stars in $VJHK$ to the 
integrated light of SSPs with $\log{t}=7.5$ (left) and $\log{t}=9.0$ 
(right). The total number of post MS stars is normalized to 100. 
The results for different IMF slopes are presented. 
The MS spectral types are indicated in the bottom panels. \label{f4}}

\figcaption[fig5.ps]{The integrated color ($J-K$) as a function of 
SSP age.
The IMF slope is $x=1.35$. In (a) it is shown the color evolution 
of a massive SSP with $10^6$ stars and a less populous SSP 
with $10^3$ stars. The numbers below the fluctuation bars indicate
the percentage of post MS stars (=Npms/(100.N)). In (b) different
simulation sets are presented with the number of post MS
fixed in 100 and 500. \label{f5}}

\figcaption[fig6.ps]{The distribution of the relative $VJHK$ 
flux contribution
with age for upper MS spectral types and post MS stars. $x=1.35$ and
Npms=100. \label{f6}}

\figcaption[fig7.ps]{(a) The $\log{t}=7.5$ and $\log{t}=9.0$  SSP 
integrated 
colors with different number of post MS stars are compared to 
integrated colors of Large Magellanic Cloud star clusters, as 
indicated by their SWB types; the larger error bars in both SSPs
correspond to the case where there is 1 post MS star, while the
smaller ones to the case where 100 post MS stars are present in
the system. (b) SSP integrated colors after inclusion of TP-AGB
for the same ages as in (a); the bluer points at both ages 
correspond to systems with total $F_{\rm V}=10$ while the redder
ones to those with total $F_{\rm V}=1000$.
 \label{f7}}

\newpage

\begin{deluxetable}{lrrrrrrrrrrr}
\tablecolumns{12}
\tablewidth{0pt}
\tablecaption{$\log{t}=7.5$}
\tablehead{
\colhead{} & \multicolumn{3}{c} {$F_{\rm V}=10$} & \colhead{} &
 \multicolumn{3} {c}
 {$F_{\rm V}=100$} & \colhead{} & \multicolumn{3} {c} {$F_{\rm V}=1000$} \\
\cline{2-4} \cline{6-8} \cline{10-12} \\
\colhead{} & \multicolumn{3}{c} {$x$(IMF)} & \colhead{} & \multicolumn{3}{c}
 {$x$(IMF)} & \colhead{} & \multicolumn{3}{c} {$x$(IMF)} \\
\colhead{} & \colhead{-0.5} & \colhead{1.35} & \colhead{2.0} & 
 \colhead{} & \colhead{-0.5} & \colhead{1.35} & \colhead{2.0} & 
 \colhead{} & \colhead{-0.5} & \colhead{1.35} & \colhead{2.0}   } 
\startdata
$K$                & -4.567 & -5.038 & -4.423 &  & -7.120 & -7.038 & -6.540 &  & -9.542 & -9.430 & -9.201 \nl
$\sigma_{\rm K}$   & 4.091 & 3.741 & 2.575 &  & 2.944 & 2.494 & 1.880 &  & 0.659 & 0.724 & 0.774 \nl\nl
$V-K$              & -0.467 & -0.080 & 0.032 &  & 0.534 & 0.527 & 0.334 &  & 1.920 & 1.680 & 1.368 \nl
$\sigma_{\rm V-K}$ & 0.894 & 0.959 & 0.743 &  & 1.705 & 1.500 & 1.078 &  & 0.578 & 0.594 & 0.653 \nl\nl
$J-K$              & -0.114 & 0.015 & 0.064 &  & 0.192 & 0.213 & 0.174 &  & 0.731 & 0.693 & 0.595 \nl
$\sigma_{\rm J-K}$ & 0.235 & 0.245 & 0.187 &  & 0.493 & 0.435 & 0.324 &  & 0.168 & 0.181 & 0.228 \nl\nl
$H-K$              & -0.055 & -0.018 & -0.005 &  & 0.017 & 0.026 & 0.021 &  & 0.153 & 0.146 & 0.125 \nl
$\sigma_{\rm H-K}$ & 0.061 & 0.059 & 0.044 &  & 0.120 & 0.101 & 0.074 &  & 0.036 & 0.039 & 0.053 \nl
\enddata
\end{deluxetable}

\newpage

\begin{deluxetable}{lrrrrrrrrrrr}
\tablecolumns{12}
\tablewidth{0pt}
\tablecaption{$\log{t}=7.75$}
\tablehead{
\colhead{} & \multicolumn{3}{c} {$F_{\rm V}=10$} & \colhead{} &
 \multicolumn{3} {c}
 {$F_{\rm V}=100$} & \colhead{} & \multicolumn{3} {c} {$F_{\rm V}=1000$} \\
\cline{2-4} \cline{6-8} \cline{10-12} \\
\colhead{} & \multicolumn{3}{c} {$x$(IMF)} & \colhead{} & \multicolumn{3}{c}
 {$x$(IMF)} & \colhead{} & \multicolumn{3}{c} {$x$(IMF)} \\
\colhead{} & \colhead{-0.5} & \colhead{1.35} & \colhead{2.0} & 
 \colhead{} & \colhead{-0.5} & \colhead{1.35} & \colhead{2.0} & 
 \colhead{} & \colhead{-0.5} & \colhead{1.35} & \colhead{2.0}   } 
\startdata
$K$  &  -4.422 &  -5.387 &  -4.919 & &  -7.817 &  -7.371 &  -6.912 & &  -9.930 &  -9.635 &  -9.334 \nl
$\sigma_{\rm K}$ &   3.603 &   3.617 &   2.786 & &   2.356 &   2.034 &   1.769 & &   0.729 &   0.761 &   0.789 \nl\nl
$V-K$  &  -0.055 &   0.004 &   0.214 & &   1.521 &   1.108 &   0.845 & &   2.327 &   1.874 &   1.551 \nl
$\sigma_{\rm V-K}$&   1.040 &   0.966 &   0.840 & &   1.557 &   1.323 &   1.155 & &   0.674 &   0.690 &   0.692 \nl\nl
$J-K$&   0.000 &   0.040 &   0.107 & &   0.533 &   0.420 &   0.330 & &   0.909 &   0.784 &   0.677 \nl
$\sigma_{\rm J-K}$&   0.281 &   0.260 &   0.202 & &   0.500 &   0.432 &   0.362 & &   0.301 &   0.304 &   0.312 \nl\nl
$H-K$   &  -0.022 &  -0.008 &   0.007 & &   0.117 &   0.088 &   0.063 & &   0.236 &   0.196 &   0.166 \nl
$\sigma_{\rm H-K}$ &   0.069 &   0.073 &   0.048 & &   0.155 &   0.127 &   0.096 & &   0.131 &   0.123 &   0.120 \nl
\enddata
\end{deluxetable}

\newpage
\begin{deluxetable}{lrrrrrrrrrrr}
\tablecolumns{12}
\tablewidth{0pt}
\tablecaption{$\log{t}=8$}
\tablehead{
\colhead{} & \multicolumn{3}{c} {$F_{\rm V}=10$} & \colhead{} &
 \multicolumn{3} {c}
 {$F_{\rm V}=100$} & \colhead{} & \multicolumn{3} {c} {$F_{\rm V}=1000$} \\
\cline{2-4} \cline{6-8} \cline{10-12} \\
\colhead{} & \multicolumn{3}{c} {$x$(IMF)} & \colhead{} & \multicolumn{3}{c}
 {$x$(IMF)} & \colhead{} & \multicolumn{3}{c} {$x$(IMF)} \\
\colhead{} & \colhead{-0.5} & \colhead{1.35} & \colhead{2.0} & 
 \colhead{} & \colhead{-0.5} & \colhead{1.35} & \colhead{2.0} & 
 \colhead{} & \colhead{-0.5} & \colhead{1.35} & \colhead{2.0}   } 
\startdata
$K$ &  -6.133 &  -4.473 &  -3.985 & &  -7.865 &  -7.154 &  -6.986 & & -10.490 & -10.069 &  -9.718 \nl
$\sigma_{\rm K}$   &   3.986 &   2.263 &   1.703 & &   1.545 &   1.164 &   1.229 & &   0.772 &   0.811 &   0.832 \nl\nl
$V-K$ &   0.625 &   0.416 &   0.496 & &   1.902 &   1.488 &   1.249 & &   2.638 &   2.197 &   1.886 \nl
$\sigma_{\rm V-K}$&   1.555 &   1.077 &   0.833 & &   1.038 &   0.844 &   0.815 & &   0.708 &   0.742 &   0.757 \nl\nl
$J-K$ &   0.196 &   0.155 &   0.187 & &   0.693 &   0.569 &   0.496 & &   1.015 &   0.899 &   0.787 \nl
$\sigma_{\rm J-K}$ &   0.430 &   0.303 &   0.239 & &   0.329 &   0.274 &   0.294 & &   0.326 &   0.337 &   0.349 \nl\nl
$H-K$    &   0.021 &   0.015 &   0.023 & &   0.133 &   0.100 &   0.089 & &   0.267 &   0.228 &   0.191 \nl
$\sigma_{\rm H-K}$&   0.106 &   0.058 &   0.044 & &   0.123 &   0.083 &   0.098 & &   0.162 &   0.156 &   0.157 \nl
\enddata
\end{deluxetable}
\newpage
\begin{deluxetable}{lrrrrrrrrrrr}
\tablecolumns{12}
\tablewidth{0pt}
\tablecaption{$\log{t}=8.25$}
\tablehead{
\colhead{} & \multicolumn{3}{c} {$F_{\rm V}=10$} & \colhead{} &
 \multicolumn{3} {c}
 {$F_{\rm V}=100$} & \colhead{} & \multicolumn{3} {c} {$F_{\rm V}=1000$} \\
\cline{2-4} \cline{6-8} \cline{10-12} \\
\colhead{} & \multicolumn{3}{c} {$x$(IMF)} & \colhead{} & \multicolumn{3}{c}
 {$x$(IMF)} & \colhead{} & \multicolumn{3}{c} {$x$(IMF)} \\
\colhead{} & \colhead{-0.5} & \colhead{1.35} & \colhead{2.0} & 
 \colhead{} & \colhead{-0.5} & \colhead{1.35} & \colhead{2.0} & 
 \colhead{} & \colhead{-0.5} & \colhead{1.35} & \colhead{2.0}   } 
\startdata
$K$ &  -4.293 &  -5.112 &  -4.011 & &  -7.975 &  -7.520 &  -7.322 & & -10.616 & -10.252 &  -9.907 \nl
$\sigma_{\rm K}$&   1.875 &   2.093 &   1.316 & &   1.585 &   1.280 &   1.225 & &   0.894 &   0.862 &   0.930 \nl\nl
$V-K$ &   0.928 &   0.913 &   0.816 & &   1.956 &   1.552 &   1.416 & &   2.830 &   2.254 &   1.926 \nl
$\sigma_{\rm V-K}$&   1.098 &   1.019 &   0.801 & &   1.057 &   0.917 &   0.822 & &   0.818 &   0.795 &   0.829 \nl\nl
$J-K$&   0.312 &   0.311 &   0.284 & &   0.697 &   0.579 &   0.535 & &   1.085 &   0.912 &   0.790 \nl
$\sigma_{\rm J-K}$&   0.341 &   0.289 &   0.248 & &   0.313 &   0.290 &   0.278 & &   0.337 &   0.351 &   0.374 \nl\nl
$H-K$  &   0.044 &   0.048 &   0.044 & &   0.138 &   0.110 &   0.101 & &   0.301 &   0.239 &   0.200 \nl
$\sigma_{\rm H-K}$&   0.061 &   0.056 &   0.044 & &   0.122 &   0.104 &   0.104 & &   0.155 &   0.153 &   0.160 \nl
\enddata
\end{deluxetable}
\newpage
\begin{deluxetable}{lrrrrrrrrrrr}
\tablecolumns{12}
\tablewidth{0pt}
\tablecaption{$\log{t}=8.5$}
\tablehead{
\colhead{} & \multicolumn{3}{c} {$F_{\rm V}=10$} & \colhead{} &
 \multicolumn{3} {c}
 {$F_{\rm V}=100$} & \colhead{} & \multicolumn{3} {c} {$F_{\rm V}=1000$} \\
\cline{2-4} \cline{6-8} \cline{10-12} \\
\colhead{} & \multicolumn{3}{c} {$x$(IMF)} & \colhead{} & \multicolumn{3}{c}
 {$x$(IMF)} & \colhead{} & \multicolumn{3}{c} {$x$(IMF)} \\
\colhead{} & \colhead{-0.5} & \colhead{1.35} & \colhead{2.0} & 
 \colhead{} & \colhead{-0.5} & \colhead{1.35} & \colhead{2.0} & 
 \colhead{} & \colhead{-0.5} & \colhead{1.35} & \colhead{2.0}   } 
\startdata
$K$&  -6.886 &  -5.978 &  -4.502 & &  -8.477 &  -8.137 &  -7.658 & & -10.977 & -10.572 & -10.335 \nl
$\sigma_{\rm K}$&   3.303 &   2.694 &   1.424 & &   1.869 &   1.907 &   1.593 & &   0.804 &   0.889 &   0.964 \nl\nl
$V-K$ &   1.545 &   1.241 &   1.131 & &   2.275 &   1.852 &   1.630 & &   3.275 &   2.848 &   2.560 \nl
$\sigma_{\rm V-K}$&   1.337 &   1.131 &   0.813 & &   1.310 &   1.281 &   1.085 & &   0.752 &   0.824 &   0.879 \nl\nl
$J-K$&   0.496 &   0.392 &   0.365 & &   0.766 &   0.631 &   0.561 & &   1.262 &   1.141 &   1.028 \nl
$\sigma_{\rm J-K}$&   0.340 &   0.275 &   0.228 & &   0.410 &   0.381 &   0.327 & &   0.311 &   0.353 &   0.382 \nl\nl
$H-K$ &   0.092 &   0.068 &   0.061 & &   0.171 &   0.135 &   0.114 & &   0.386 &   0.342 &   0.297 \nl
$\sigma_{\rm H-K}$&   0.101 &   0.063 &   0.043 & &   0.163 &   0.138 &   0.110 & &   0.143 &   0.153 &   0.161 \nl
\enddata
\end{deluxetable}
\newpage
\begin{deluxetable}{lrrrrrrrrrrr}
\tablecolumns{12}
\tablewidth{0pt}
\tablecaption{$\log{t}=8.75$}
\tablehead{
\colhead{} & \multicolumn{3}{c} {$F_{\rm V}=10$} & \colhead{} &
 \multicolumn{3} {c}
 {$F_{\rm V}=100$} & \colhead{} & \multicolumn{3} {c} {$F_{\rm V}=1000$} \\
\cline{2-4} \cline{6-8} \cline{10-12} \\
\colhead{} & \multicolumn{3}{c} {$x$(IMF)} & \colhead{} & \multicolumn{3}{c}
 {$x$(IMF)} & \colhead{} & \multicolumn{3}{c} {$x$(IMF)} \\
\colhead{} & \colhead{-0.5} & \colhead{1.35} & \colhead{2.0} & 
 \colhead{} & \colhead{-0.5} & \colhead{1.35} & \colhead{2.0} & 
 \colhead{} & \colhead{-0.5} & \colhead{1.35} & \colhead{2.0}   } 
\startdata
$K$&  -6.173 &  -5.472 &  -5.067 & &  -8.541 &  -8.258 &  -8.222 & & -11.186 & -10.773 & -10.703 \nl
$\sigma_{\rm K}$&   2.369 &   1.890 &   1.449 & &   1.702 &   1.703 &   1.627 & &   0.755 &   0.870 &   0.841 \nl\nl
$V-K$&   1.626 &   1.429 &   1.418 & &   2.416 &   2.087 &   1.985 & &   3.398 &   2.932 &   2.773 \nl
$\sigma_{\rm V-K}$&   0.946 &   0.831 &   0.794 & &   1.221 &   1.207 &   1.144 & &   0.695 &   0.799 &   0.776 \nl\nl
$J-K$&   0.503 &   0.439 &   0.431 & &   0.793 &   0.688 &   0.666 & &   1.283 &   1.147 &   1.112 \nl
$\sigma_{\rm J-K}$&   0.268 &   0.213 &   0.199 & &   0.413 &   0.416 &   0.418 & &   0.292 &   0.362 &   0.374 \nl\nl
$H-K$&   0.090 &   0.076 &   0.074 & &   0.183 &   0.155 &   0.151 & &   0.400 &   0.352 &   0.345 \nl
$\sigma_{\rm H-K}$&   0.082 &   0.055 &   0.045 & &   0.171 &   0.166 &   0.166 & &   0.138 &   0.164 &   0.173 \nl
\enddata
\end{deluxetable}
\newpage
\begin{deluxetable}{lrrrrrrrrrrr}
\tablecolumns{12}
\tablewidth{0pt}
\tablecaption{$\log{t}=9$}
\tablehead{
\colhead{} & \multicolumn{3}{c} {$F_{\rm V}=10$} & \colhead{} &
 \multicolumn{3} {c}
 {$F_{\rm V}=100$} & \colhead{} & \multicolumn{3} {c} {$F_{\rm V}=1000$} \\
\cline{2-4} \cline{6-8} \cline{10-12} \\
\colhead{} & \multicolumn{3}{c} {$x$(IMF)} & \colhead{} & \multicolumn{3}{c}
 {$x$(IMF)} & \colhead{} & \multicolumn{3}{c} {$x$(IMF)} \\
\colhead{} & \colhead{-0.5} & \colhead{1.35} & \colhead{2.0} & 
 \colhead{} & \colhead{-0.5} & \colhead{1.35} & \colhead{2.0} & 
 \colhead{} & \colhead{-0.5} & \colhead{1.35} & \colhead{2.0}   } 
\startdata
$K$&  -5.613 &  -4.873 &  -5.719 & &  -8.386 &  -8.041 &  -8.165 & & -10.863 & -10.527 & -10.581 \nl
$\sigma_{\rm K}$&   1.572 &   1.104 &   1.573 & &   1.447 &   1.329 &   1.311 & &   0.680 &   0.719 &   0.712 \nl\nl
$V-K$&   1.895 &   1.719 &   1.709 & &   2.484 &   2.198 &   2.161 & &   3.122 &   2.801 &   2.668 \nl
$\sigma_{\rm V-K}$&   0.749 &   0.680 &   0.638 & &   1.095 &   0.976 &   0.990 & &   0.633 &   0.668 &   0.665 \nl\nl
$J-K$&   0.559 &   0.503 &   0.510 & &   0.772 &   0.695 &   0.689 & &   1.132 &   1.038 &   0.998 \nl
$\sigma_{\rm J-K}$&   0.188 &   0.170 &   0.195 & &   0.371 &   0.352 &   0.370 & &   0.277 &   0.317 &   0.328 \nl\nl
$H-K$&   0.096 &   0.085 &   0.089 & &   0.171 &   0.149 &   0.152 & &   0.334 &   0.303 &   0.291 \nl
$\sigma_{\rm H-K}$&   0.058 &   0.040 &   0.065 & &   0.151 &   0.143 &   0.150 & &   0.127 &   0.145 &   0.149 \nl
\enddata
\end{deluxetable}
\newpage
\begin{deluxetable}{lrrrrrrrrrrr}
\tablecolumns{12}
\tablewidth{0pt}
\tablecaption{$\log{t}=9.25$}
\tablehead{
\colhead{} & \multicolumn{3}{c} {$F_{\rm V}=10$} & \colhead{} &
 \multicolumn{3} {c}
 {$F_{\rm V}=100$} & \colhead{} & \multicolumn{3} {c} {$F_{\rm V}=1000$} \\
\cline{2-4} \cline{6-8} \cline{10-12} \\
\colhead{} & \multicolumn{3}{c} {$x$(IMF)} & \colhead{} & \multicolumn{3}{c}
 {$x$(IMF)} & \colhead{} & \multicolumn{3}{c} {$x$(IMF)} \\
\colhead{} & \colhead{-0.5} & \colhead{1.35} & \colhead{2.0} & 
 \colhead{} & \colhead{-0.5} & \colhead{1.35} & \colhead{2.0} & 
 \colhead{} & \colhead{-0.5} & \colhead{1.35} & \colhead{2.0}   } 
\startdata
$K$&  -4.755 &  -4.851 &  -4.684 & &  -7.350 &  -7.394 &  -7.252 & &  -9.816 &  -9.863 &  -9.719 \nl
$\sigma_{\rm K}$&   0.671 &   0.562 &   0.603 & &   0.224 &   0.208 &   0.218 & &   0.081 &   0.074 &   0.077 \nl\nl
$V-K$&   2.149 &   2.038 &   1.942 & &   2.332 &   2.171 &   2.108 & &   2.325 &   2.165 &   2.102 \nl
$\sigma_{\rm V-K}$&   0.430 &   0.373 &   0.402 & &   0.165 &   0.157 &   0.167 & &   0.060 &   0.057 &   0.060 \nl\nl
$J-K$&   0.634 &   0.603 &   0.570 & &   0.701 &   0.657 &   0.638 & &   0.701 &   0.658 &   0.639 \nl
$\sigma_{\rm J-K}$&   0.128 &   0.117 &   0.130 & &   0.054 &   0.054 &   0.059 & &   0.020 &   0.020 &   0.022 \nl\nl
$H-K$&   0.106 &   0.101 &   0.094 & &   0.121 &   0.113 &   0.110 & &   0.121 &   0.114 &   0.110 \nl
$\sigma_{\rm H-K}$&   0.028 &   0.025 &   0.027 & &   0.014 &   0.014 &   0.015 & &   0.005 &   0.005 &   0.006 \nl
\enddata
\end{deluxetable}


\begin{thebibliography}{}
 \bibitem[Allen et al.\ (1996)]{all96} Allen, R. C., Bottcher, C., Bording, P.,
          Burns, P., Conery, J., Davies, T. R., Demmel, J., Johnson, C., 
          Kantha, L., Martin, W., Parks, G., Piacsek, S., Pryor, D., 
          Schlick, T., Strayer, M. R., Umar, V. M., Voigt, R., Wagener, J.,
          Zachmann, D., \& Ziebarth, J.  1996, Computacional Science
          Education Project, U. S. Department of Energy, 
          http://csep1.phy.ornl.gov/CSEP/MC/MC.html
 \bibitem[Barbaro \& Bertelli (1977)]{bar77} Barbaro, G., \&
          Bertelli, G.  1977, \aap, 54, 243
 \bibitem[Becker \& Mathews (1983)]{bec83} Becker, S. A., \&
          Mathews, G. J.  1983, \apj, 270, 155
 \bibitem[Bedijn (1988)]{bed88} Bedijn, P. J. 1988, \aap, 205, 105
 \bibitem[Bertelli et al.\ (1994)]{ber94} Bertelli, G., Bressan, A., 
          Chiosi, C., Fagotto, F., \& Nasi, E.  1994, \aaps, 106, 275
 \bibitem[Bessell (1991)]{bes91} Bessell, M. S.  1991, \aj, 101, 662 (B91)
 \bibitem[Bessell \& Brett (1988)]{bes88} Bessell, M. S., \& 
          Brett, J. M.  1988, \pasp, 100, 1134
 \bibitem[Bica et al.\ 1996]{bic96} Bica, E., Clari\'a, J. J., 
          Dottori, H., Santos Jr., J. F. C., \&
          Piatti, A. E.  1996, \apjs, 102, 57
 \bibitem[Charlot \& Bruzual (1991)]{cha91} Charlot, S., \&
          Bruzual, G. A.  1991, \apj 367, 126
 \bibitem[Chiosi et al.\ (1988)]{chi88} Chiosi, C., Bertelli, G., 
          \& Bressan, A. 1988, \aap, 196, 84
 \bibitem[Ferraro et al.\ (1995)]{fer95} Ferraro, F. R., Fusi Pecci, F., 
          Testa, V., Greggio, L., Corsi, C. E., Buonanno, R., 
          Terndrup, D. M., \& Zinnecker, H. 1995, \mnras, 272, 391
 \bibitem[Frogel 1988]{fro88} Frogel, J. A. 1988, \araa, 26, 51
 \bibitem[Frogel et al.\ 1990]{fro90} Frogel, J. A., Mould, J., \& 
          Blanco, V. M.  1990, \apj, 352, 96
 \bibitem[Frogel et al.\ (1980)]{fro80} Frogel, J. A., Persson, S. E.,
          \& Cohen, J. G. 1980, \apj, 239, 495
 \bibitem[Frogel \& Whitford (1987)]{fro87} Frogel, J. A., \&
          Whitford, A. E. 1987, \apj, 320, 199
 \bibitem[Girardi \& Bica (1993)]{gir93} Girardi, L., 
          \& Bica, E.  1993, \aap, 274, 279
 \bibitem[Girardi et al.\ (1995)]{gir95} Girardi, L., Chiosi, C., 
          Bertelli, G., \& Bressan, A. 1995, \aap 298, 87 
 \bibitem[Koornneef (1983)]{koo83} Koornneef, J.  1983, \aap, 128, 84
 \bibitem[Maeder \& Meynet (1989)]{mae89} Maeder, A., \&
          Meynet, G.  1989, \aap, 210, 155
 \bibitem[M\'era et al.\ 1996]{mer96} M\'era, D., Chabrier, G., 
          \& Baraffe, I.  1996, \apj, 459, L87
 \bibitem[Persson et al.\ 1983]{per83} Persson, S. E., Aaronson, M., 
          Cohen, J. G., Frogel, J. A., \& Matthews, K.  1983, \apj, 266, 105
 \bibitem[Press et al.\ (1992)]{pre92} Press, W. H., Teukolsky, S. A., 
          Vetterling, W. T., \& Flannery, B. P.  1992, Numerical Recipes 
          in Fortran, 2nd edition, Cambridge University Press
 \bibitem[Salpeter (1955)]{sal55} Salpeter, E. E.  1955, \apj, 121, 161
 \bibitem[Schaller et al.\ 1992]{sch92} Schaller, G., Schaerer, D., 
         Meynet, G., \& Maeder, A.  1992, \aaps, 96, 269 (SSMM)
 \bibitem[Schmidt-Kaler 1982]{sch82} Schmidt-Kaler, T.  1982, in:
          Landolt-B\"ornstein, Stars and Star Clusters, ed. K. Schaifers 
          and H. H. Voigt, Springer-Verlag, Berlin, vol. 2b, p. 1 (SK82)
 \bibitem[VandenBerg et al.\ (1983)]{van83} VandenBerg, D. A., 
          Hartwick, F. D. A., Dawson, P., \& Alexander, D. R.  
          1983, \apj, 266, 747
 \end{thebibliography}
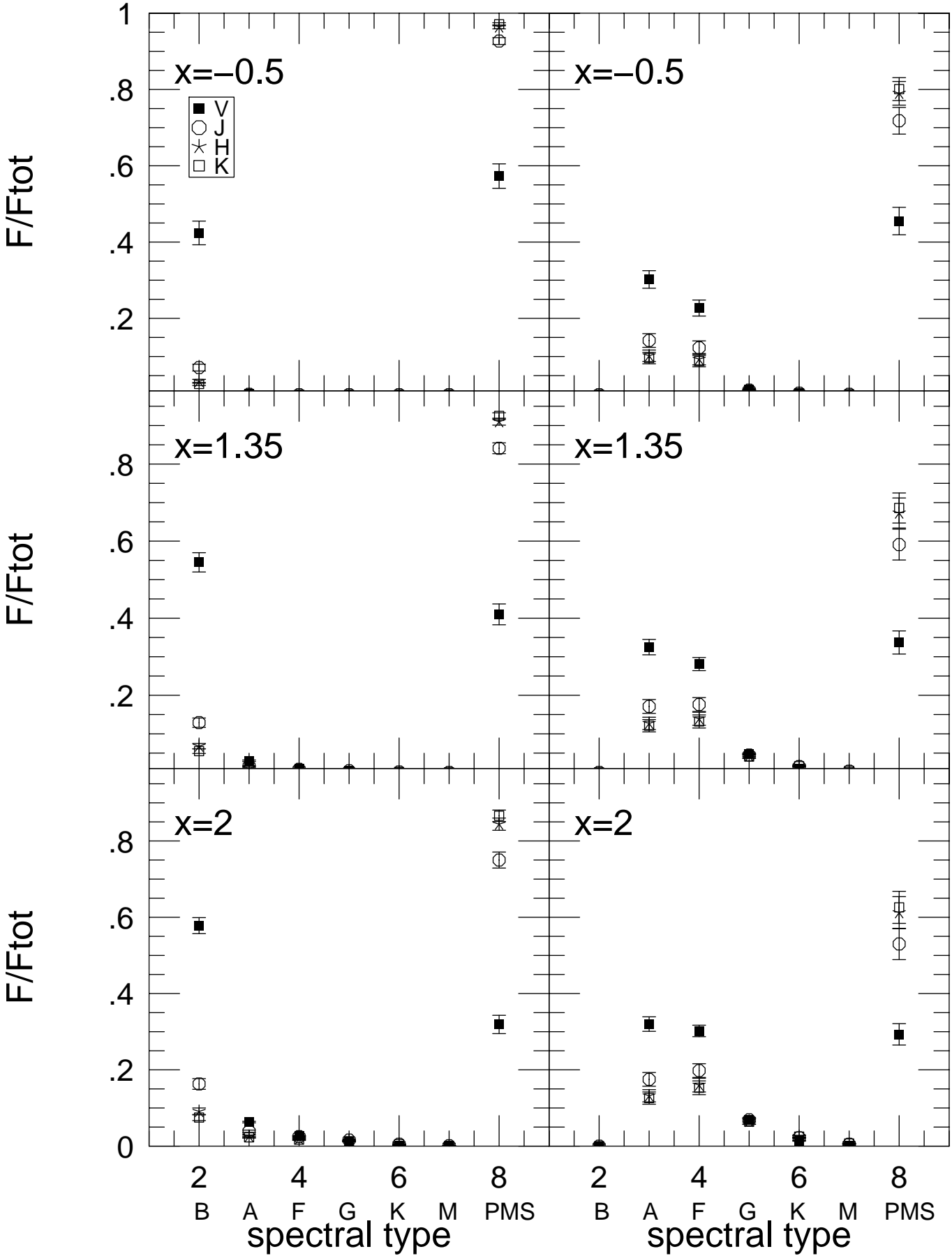
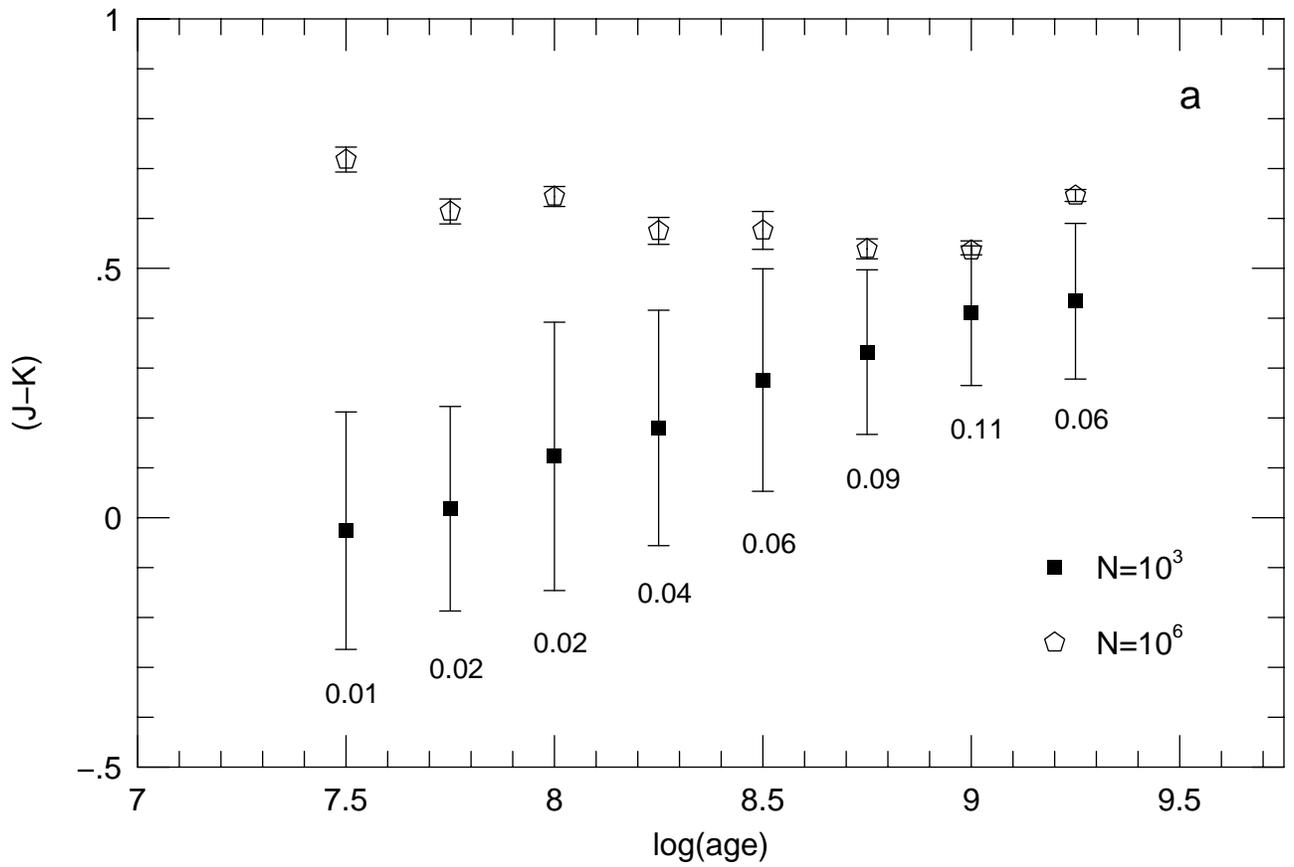
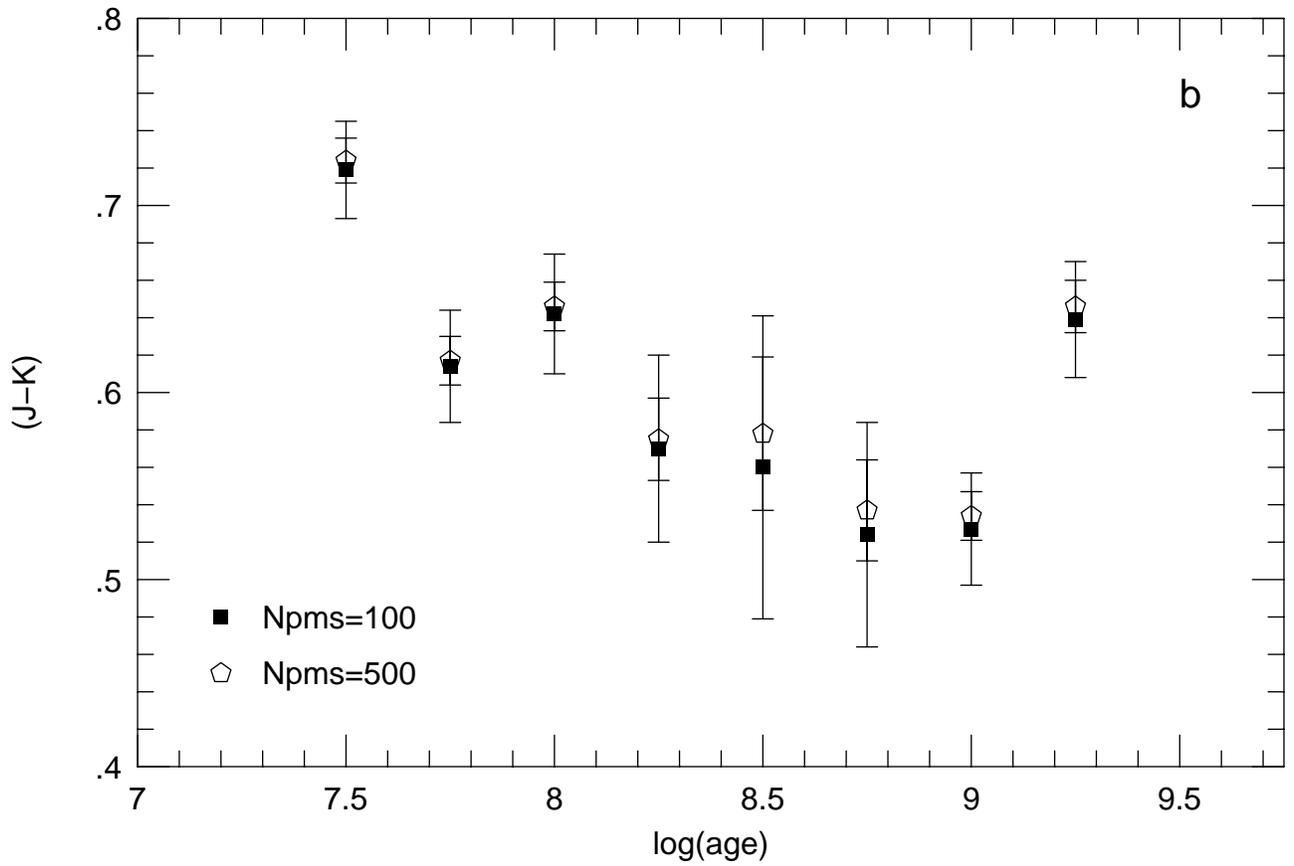
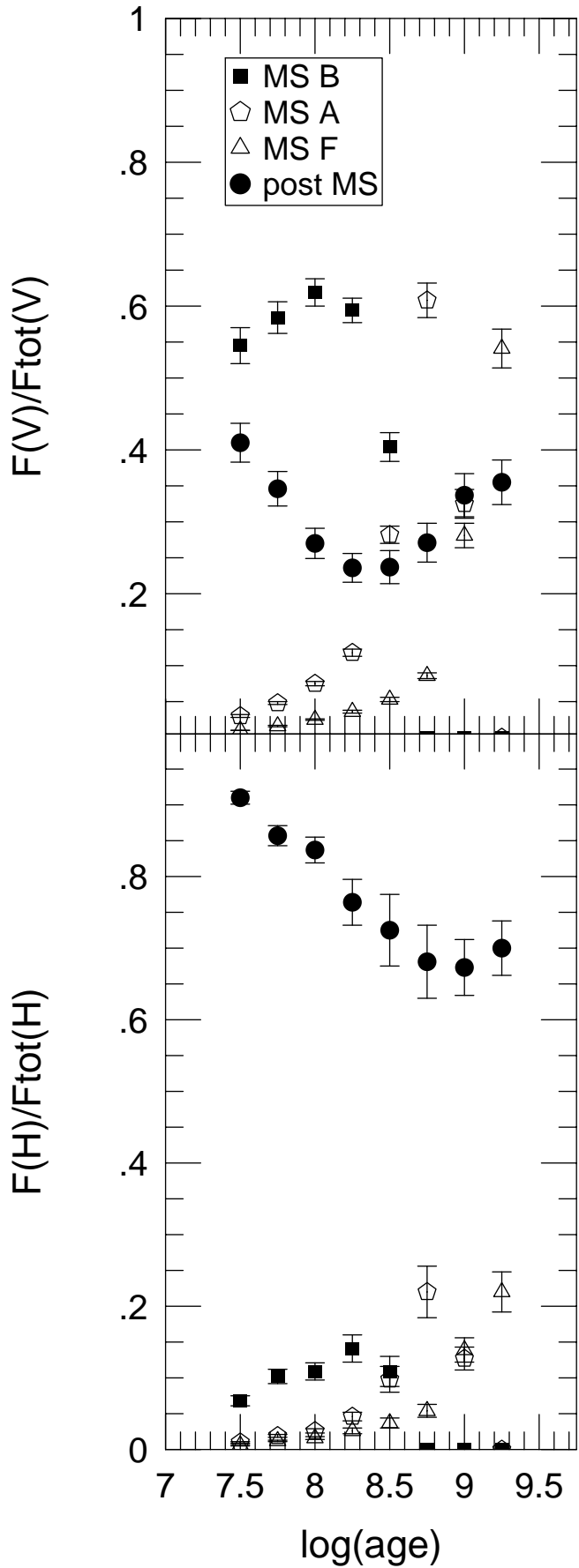
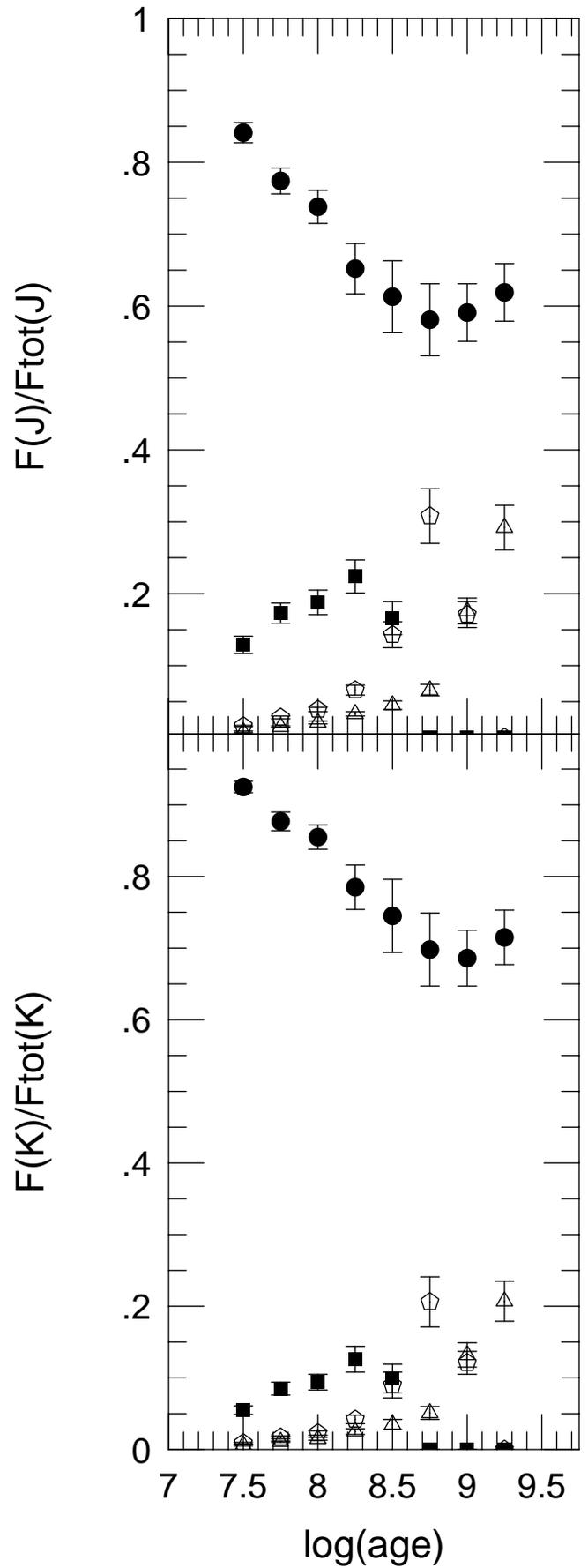
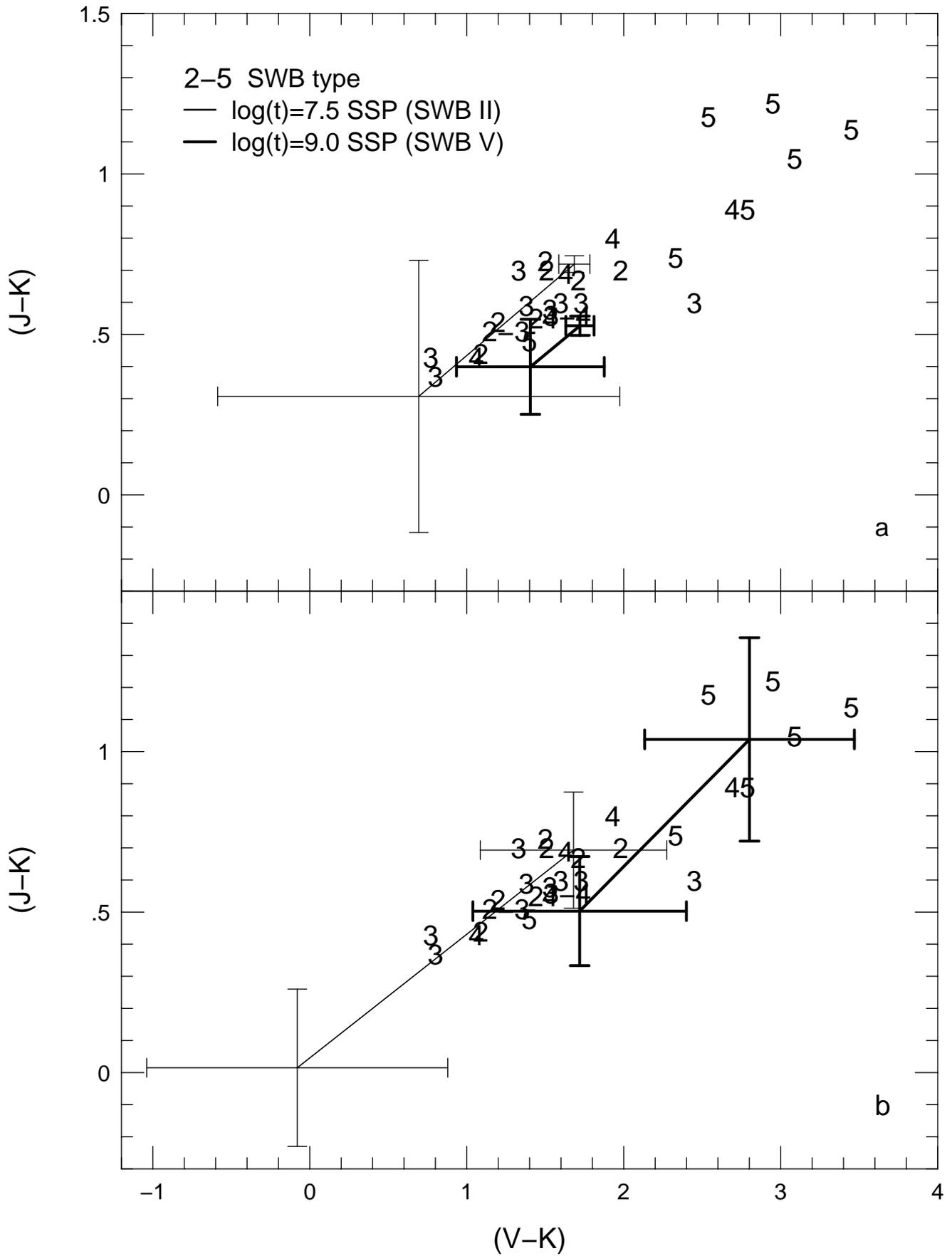
\end{document}